\documentclass[a4paper,runningheads]{llncs}

\usepackage{amsmath}
\usepackage{graphicx}
\usepackage{subfigure}

\newcommand{\cciteseer}{\operatorname{complete\_citeseer}}
\newcommand{\cdblp}{\operatorname{complete\_dblp}}
\newcommand{\citeseer}{\operatorname{citeseer}}
\newcommand{\dblp}{\operatorname{dblp}}
\newcommand{\all}{\operatorname{all}}
\newcommand{\pf}{\operatorname{pf}}
\newcommand{\pp}{\operatorname{pp}}
\newcommand{\p}{\operatorname{p}}
\newcommand{\ratio}{\operatorname{r}}

\newcommand{\doublefig}[4]{
\begin{figure}[ht!]
\begin{minipage}[ht]{0.45\textwidth}
\includegraphics[width=\textwidth]{#1} \caption{#2\label{fig:#1}} 
\end{minipage}
\hfill
\begin{minipage}[ht]{0.45\textwidth}
\includegraphics[width=\textwidth]{#3} \caption{#4\label{fig:#3}} 
\end{minipage}
\end{figure}
}

\begin{document}

\pagestyle{empty}

\mainmatter

\title{A Comparison of On-line Computer Science Citation Databases}

\titlerunning{Lecture Notes in Computer Science}

\author{
Vaclav Petricek\inst{1} \and Ingemar J. Cox\inst{1} \and Hui Han\inst{2} \and Isaac G. Councill\inst{3} \and C. Lee Giles\inst{3}\\
 }

\authorrunning{Vaclav Petricek et al.}

\institute{
       University College London,
       WC1E 6BT, Gower Street,
       London, United Kingdom,
\email{v.petricek@cs.ucl.ac.uk},
\email{ingemar@ieee.org}
\and
       {Yahoo! Inc.},
       {701 First Avenue},
       {Sunnyvale, CA, 94089},
\email{huihan@yahoo-inc.com}
\and
       {The School of Information Sciences and Technology},
       {The Pennsylvania State University},
       {University Park, PA 16802, USA},
\email{igc2@psu.edu},
\email{giles@ist.psu.edu}}

\maketitle

\begin{abstract}

This paper examines the difference and similarities between the two
on-line computer science citation databases DBLP and CiteSeer. The
database entries in DBLP are inserted manually while the CiteSeer
entries are obtained autonomou\-sly via a crawl of the Web and automatic
processing of user submissions.  
CiteSeer's autonomous citation database can be considered a form of
self-selected on-line survey. It is important to understand the
limitations of such databases, particularly when citation information
is used to assess the performance of authors, institutions and
funding bodies.

We show that the CiteSeer database
contains considerably fewer single author papers. This bias can be
modeled by an exponential process with intuitive explanation. The model
permits us to predict that the DBLP database covers approximately 24\%
of the entire literature of Computer Science. CiteSeer is also biased against low-cited papers. 

Despite their difference, both databases exhibit similar and
significantly different citation distributions compared with previous
analysis of the Physics community. In both databases, we also observe
that the number of authors per paper has been increasing over
time.

\end{abstract}

\section{Introduction}

Several public\footnote{
By public, we mean that access to the database
is free of charge. Commercial databases are also available, the most
well-known being the science-citation index~\cite{SCICITE}
} 
databases
of research papers became available due to the advent of the Web
\cite{arxiv,lawrence99digital,ley@DBLP,CORR,compuscience,CSbibtex,sciencedirect}
These databases collect papers in different scientific
disciplines, index them and annotate them with additional metadata. As
such, they provide an important resource for (i) finding publications,
(ii) identifying important, i.e. highly cited, papers, and (iii)
locating papers that cite a particular paper. In addition, author and document
citation rates are increasingly being used to quantify the scientific
impact of scientists, publications, journals and funding agencies.

Within the computer science community, there are two popular public
citation databases. These are DBLP~\cite{ley@DBLP} and CiteSeer
\cite{lawrence99digital}. The two databases are constructed in very
different ways. In DBLP, each entry is manually inserted by a group of
volunteers and occasionally hired students. The entries are obtained from conference proceeding and
journals. In contrast, each entry in CiteSeer is automatically entered
from an analysis of documents found on the Web. There are advantages
and disadvantages to both methods and we discuss these issues in more
detail in the next Section.

In Section~\ref{sec:datasets} we compare the two databases based on
the distribution of number of authors. We reveal that there are very pronounced
differences which appear to be primarily due to the absence of very
many single author papers in the CiteSeer database. A probabilistic model
for document acquisition is then developed that provides an intuitive
explanation for this phenomenon in Section \ref{sec:models}.

There have been a number of studies on the distribution of citations
\cite{lehmann@citation,redner@popular,laherrere@stretched,tsallis@are} and the
number of collaborators~\cite{newman@structure} using other on-line
databases. This literature is reviewed in
Section~\ref{sec:previous}. We replicate some of these studies and
show that citation distributions from both DBLP and CiteSeer differ
considerably from those reported in other research communities.

\section{The DBLP and CiteSeer databases\label{sec:datasets}}

There are a number of public, on-line computer science databases
\cite{arxiv,lawrence99digital,ley@DBLP,CORR,compuscience,CSbibtex}. The
CS BiBTeX database~\cite{CSbibtex} contains a collection of over 1.4 million
references. However, only 19,000 entries currently contain
cross-references to citing or cited publications. The Compuscience
database~\cite{compuscience} contains approximately 400,000 entries. The Computing Research Repository CoRR~\cite{CORR} contains papers from 36 areas of computer science and is now part of ArXiV~\cite{arxiv} that covers Physics, Mathematics, Nonlinear Sciences, Computer Science and Quantitative Biology. Networked Computer Science Technical Reference Library is a repository of Computer Science Technical Reports located at Old Dominion University.

DBLP was created by Michael Ley in 1998~\cite{ley@DBLP}. It currently
contains over 550,000 computer science references from around 368,000 authors.
CiteSeer was created by Steve Lawrence and C. Lee Giles in 1997
\cite{lawrence99digital}. It currently contains over 716,797 documents.

We chose to examine DBLP and CiteSeer due to the
availability of detailed citation information and their popularity.

In our analysis we focus on the difference in data acquisition and the
biases that this difference introduces.

\subsection{The differences between DBLP and CiteSeer databases}

While both the DBLP and CiteSeer databases contain computer science
bibliography and citation data, their acquisition methods greatly vary.
In this section we first discuss these differences in acquisition methods, then we look at the distribution of papers over time in each dataset, and after that we compare the distribution in the number of authors per paper.
Section~\ref{sec:models} then describes acquisition models for both DBLP and CiteSeer.

\subsubsection{Data acquisition\label{sec:acquisition}}

At the time of writing, DBLP contains over 550,000 bibliographic
entries.  Papers in DBLP originally covered database systems and logic programing. Currently DBLP also includes theory of information, automata, complexity, bioinformatics and other areas. Database entries are obtained by a limited
circle of volunteers who manually enter tables of contents of journals
and conference proceedings. The volunteers also manually entered
citation data as part of compiling the ACM anthology CD/DVDs. Corrections that are submitted to the maintainer are
also manually checked before committing.  Though the breadth of coverage
may be more narrow than CiteSeer, DBLP tries to ensure comprehensive
and complete coverage within its scope. The coverage of ACM, IEEE and LNCS is around 80--90\%. The narrower focus of DBLP is
partially enforced by the cost associated with manual entry. Although
there is the possibility of human error in the manual process of DBLP,
its metadata is generally of higher quality than
automatically extracted metadata\footnote{
This remains true, despite the recent
improvement of automatic extraction algorithms by use of support
vector machines~\cite{han@automatic}.
}.

In our analysis we used a DBLP dataset consisting of 496,125
entries. From this we extracted a dataset of
352,024 papers that specified the year of publication and the number
of authors. Only papers published between 1990 and 2002 were included,
due to the low number of papers available outside of this range.

CiteSeer currently contains over 716,797 bibliographic
entries. Automatic crawlers have the potential of achieving higher
coverage as the cost of automatic indexing is lower than for manual entry.
However, differences in typographic conventions make it hard to
automatically extract metadata such as author names, date of
publication, etc. 

CiteSeer entries can be acquired in two modes.  First, the publication
may be encountered during a crawl\footnote{CiteSeer is not performing a brute force crawl of the web but crawling a set of starting pages to the depth of 4-7}. In this case, the document will be
parsed, and title, author and other information will be entered into
the database. Second, during this parsing operation, a document's bibliography
is also analyzed and previously unknown cited documents are also entered
into the database.

CiteSeer is continuously updated with user submissions. Currently
updates are performed every two weeks. However, it
was not updated at all during the period from about March 2003 to April 2004.
Prior to March 2003 crawls were made with declining regularity.  As of
July 2004 CiteSeer has been continuously crawling the web to find new
content using user submissions, conference, and journal URLs as entry
points.

In our analysis, we used a CiteSeer dataset consisting of 575,068 entries.
From this we extracted a dataset of 325,046 papers that specified the
year of publication and the number of authors. Once again, only papers
published between 1990 and 2002 were considered. It is also important
to note that this dataset only contained entries that CiteSeer acquired by
parsing the actual document on the Web, i.e. documents that were only cited
but not actually parsed, were not included. 
We assume that the number of parsing errors is independent of the number of
authors and does not introduce any new bias.

CiteSeer may be considered a form of self-selected on-line survey - authors may choose to upload the URL where their publications are available
for subsequent crawling by CiteSeer. This self-selection introduces a bias in the CiteSeer database that
we discuss later. 
A fully automatic scientometric system
is also potentially susceptible to ``shilling'' attacks, i.e. authors
trying to alter their citation ranking by, for example,
submitting fake papers citing their work. This later issue is not
discussed further here, but appears related to similar problems
encountered by recommender systems~\cite{shyong@shilling}.

\subsubsection{Accumulation of papers per year}

In order to compare the two databases, we first examined the number of
publications in the two datasets for the years 1990 through 2002. These
years were chosen to ensure that a sufficient number of
papers per year is available in both datasets. 
\doublefig{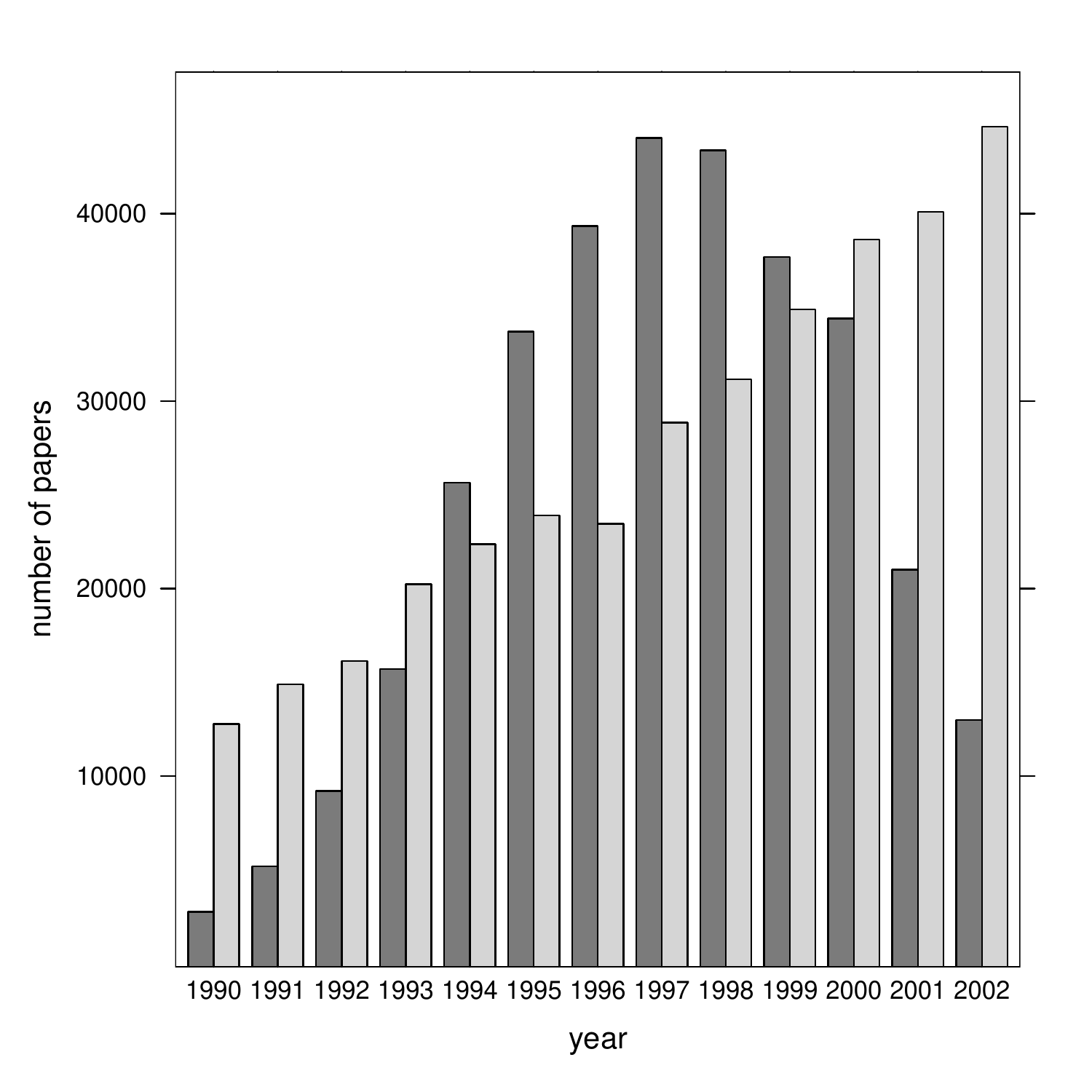}{Number of papers published in the years from 1990 to 2002 present in the DBLP (light) and CiteSeer (dark) databases.}{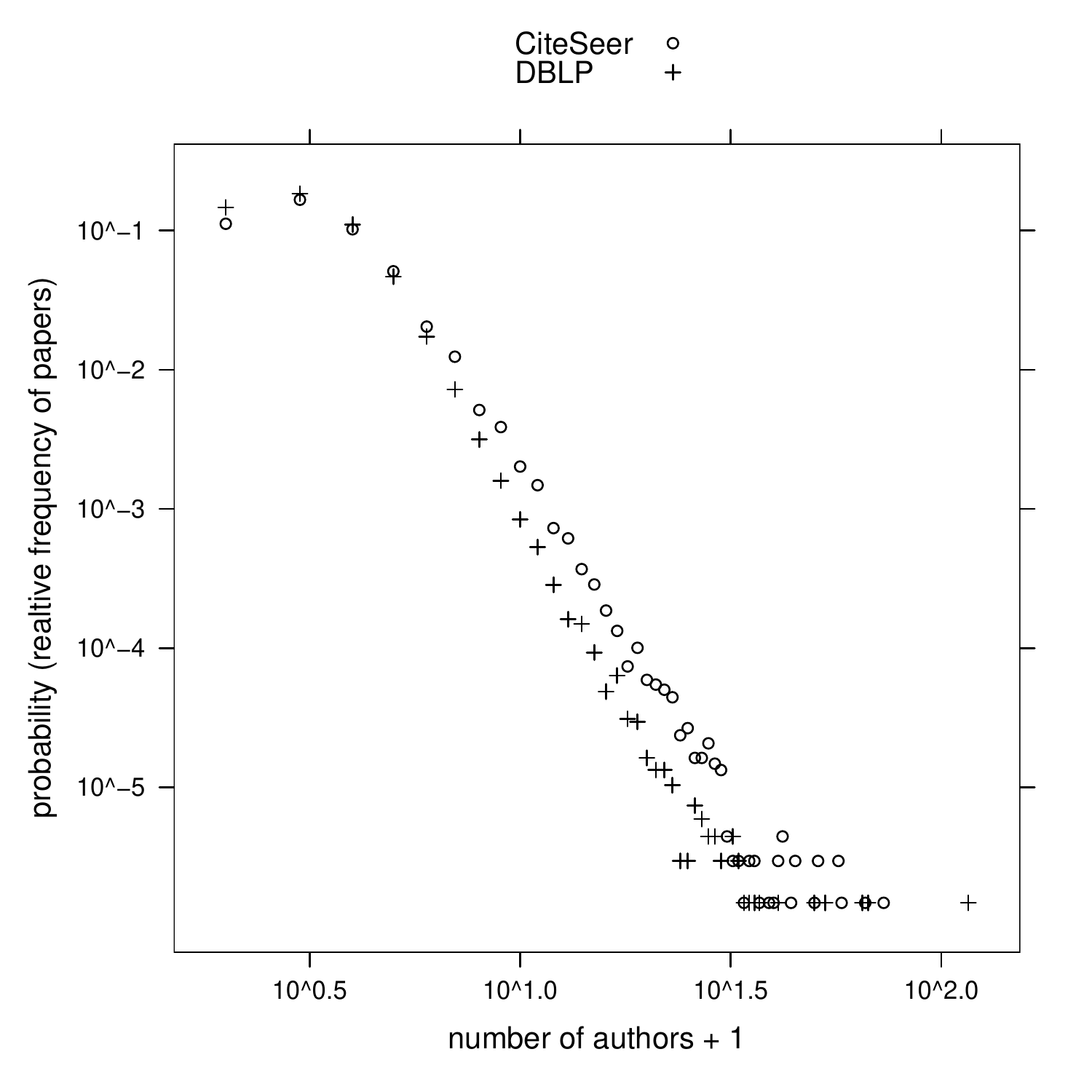}{Probability histogram of number of authors. (double logarithmic scale.)}

Figure~\ref{fig:combined-pubyear.pdf} shows a considerable difference
in the number of papers present in the two databases on an annual
basis.

The increase in the papers per year exhibited by DBLP is probably
explained by a combination of (i) the increasing number of
publications each year~\cite{price63little,ley@dblpcomputer} and (ii)
an increase in the coverage of DBLP thanks to additional funding and
improvement in processing efficiency\footnote
{
Personal communication with Michael Ley
}.

The decrease in the number of papers per year exhibited by CiteSeer
since 1997 is mainly due to (i) declining maintenance, although (ii) declining coverage (iii) intellectual property concerns (iv) dark matter effect \cite{bailey@dark} (v) end of web fever and (vi) specifics of submission process, may also have contributed.

\subsubsection{Team size}

\begin{figure}[ht]
\begin{center}
\subfigure[CiteSeer]{\includegraphics[width=.48\textwidth]{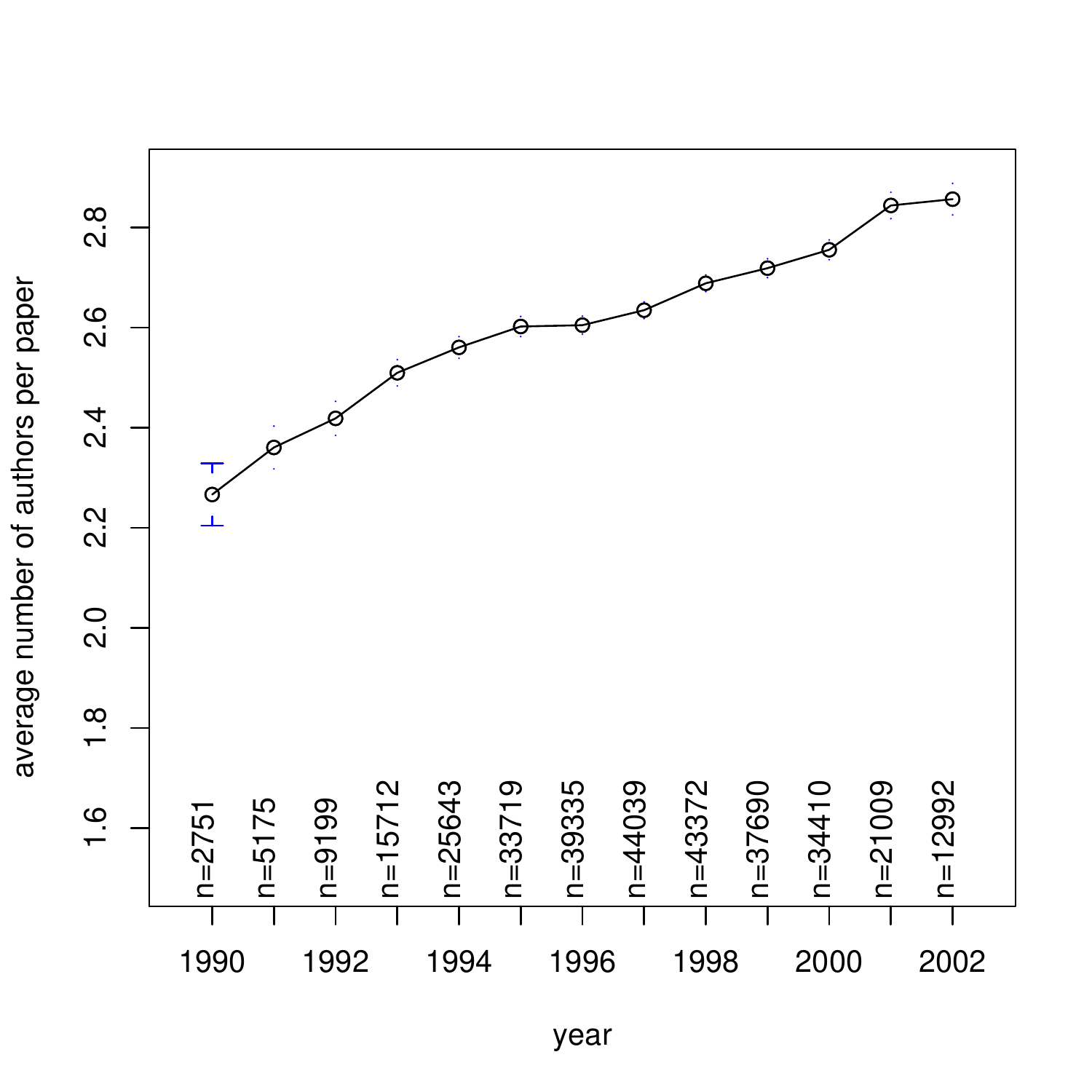}}
\subfigure[DBLP]{\includegraphics[width=.48\textwidth]{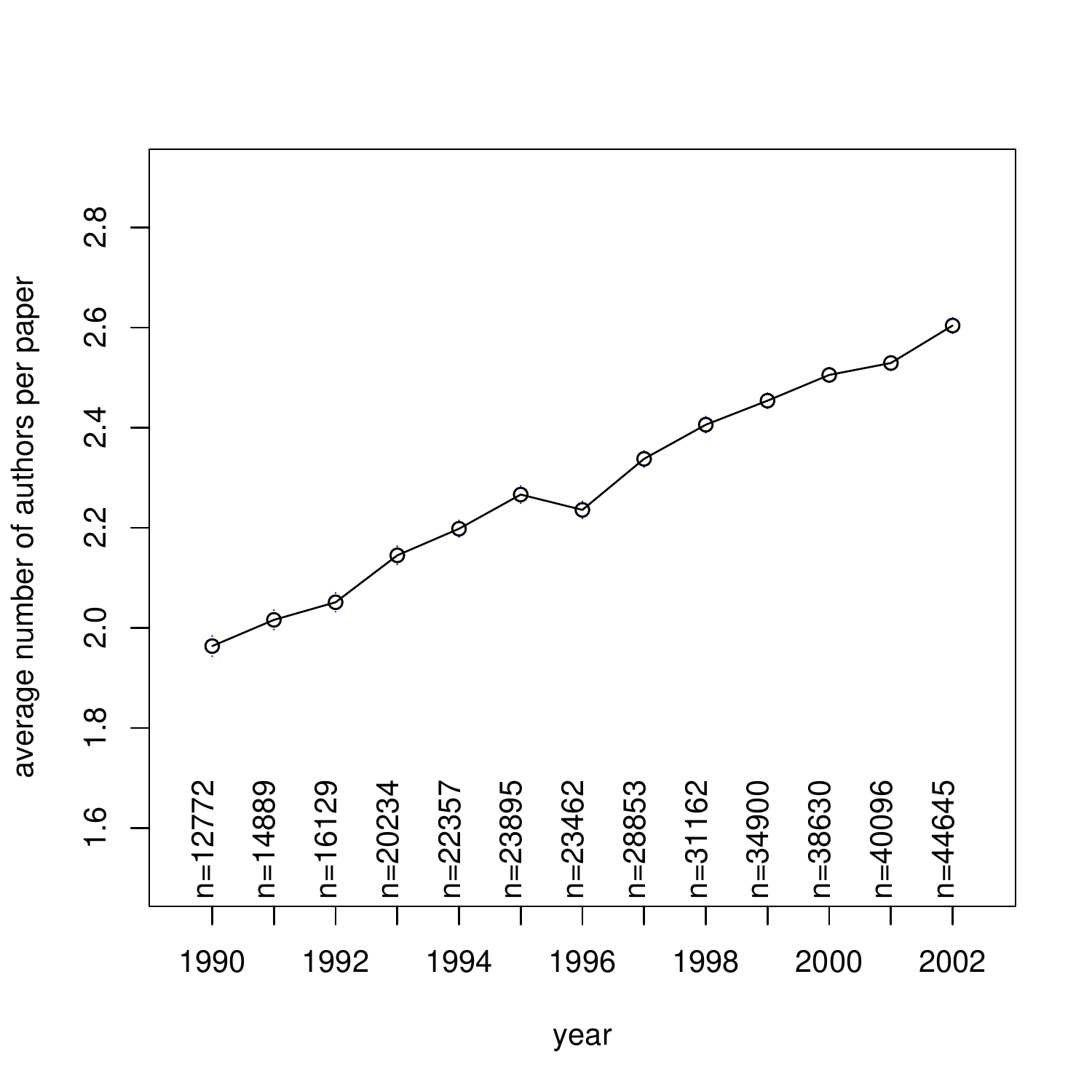}}
\label{fig:collab-authavg.pdf}
\caption{Average number of authors per paper for the years 1990 to 2002\label{fig:collab-authavg}}
\end{center}
\end{figure}

We also examined the average number of authors for papers published
between 1990 and 2002, see Figure~\ref{fig:collab-authavg}.
In both datasets, the average
is seen to be rising. It is uncertain what is causing this rise in
multi-authorship. Possible explanations include (i) funding agencies
preference to fund collaborative research
and/or (ii) collaboration has become easier with the increasing use of
email and the Web.
We observe that the CiteSeer database contains a
higher number of multi-author papers.

\subsubsection{Bias in number of authors}

Figure~\ref{fig:authprobhistogram-loglog.pdf} examines the relative
frequency of $n$-authored papers in the two datasets. Note that the
data is on a log-log scale. It is clear that CiteSeer has far fewer
single and two authored papers. In fact, CiteSeer has relatively fewer papers
published by one to three authors.  This is emphasized in
Figure~\ref{fig:authprob-dblpcsratio.pdf} in which we plot the ratio
of the frequency of $n$-authored papers in DBLP and CiteSeer for one to fifty authors. Here
we see the frequency of single-authored papers in CiteSeer is only
77\% of that ocurring in DBLP. As the number of authors increases,
the ratio decreases since CiteSeer has a higher frequency of
$n$-authored papers for $n>3$. For $n>30$, the ratio is somewhat random,
reflecting the scarcity of data in this region. We therefore limit our
analysis to numbers of authors where there are at least 100 papers in
each dataset.  This restricts the number of authors to less than 17.

As we see in Figure~\ref{fig:authprobhistogram-loglog.pdf} the number
of authors follows a power law corresponding to a line with slope
approximately $-0.23$ for DBLP and $-0.24$ for CiteSeer. 
There is an obvious cut-off from the power law for papers with low number of authors.  For CiteSeer, we hypothesize that
(i) papers with more authors are more likely to be submitted
to CiteSeer and (ii) papers with more authors appear on more homepages and are
therefore more likely to be found by the crawler. 
These ideas are modeled in Section~\ref{sec:models}.

However none of these factors is relevant to DBLP, which also exhibits
a similar drop off in single-authored papers. Other explanations may
be that (i) single author papers are less likely to be finished and
published, (ii) funding agencies encourage collaborative and
therefore multi-authored research and (iii) it is an effect of limited number of scientists in the world~\cite{laherrere@stretched}.

\subsection{DBLP and CiteSeer data acquisition models}
\label{sec:models}

To explain the apparent bias of CiteSeer towards papers with larger
numbers of authors, we develop two possible models for the acquisition of
papers within CiteSeer. We also provide a simple acquisition model for DBLP.

The first CiteSeer model is based on authors submitting their papers directly to
the database. The second CiteSeer model assumes that the papers are obtained by
a crawl of the Web. We show that in fact, both models are equivalent.

To begin, let ${\citeseer}(i)$ be the number of papers in CiteSeer with
$i$ authors, $\dblp(i)$ the number of papers in DBLP with $i$ authors
and $\all(i)$ the number of  papers with $i$ authors published in all
Computer Science. 

For DBLP, we assume a simple paper acquisition model such that there is a probability $\alpha$ that
a paper is included in DBLP and that this probability is independent
of the number of authors.  

For CiteSeer we assume that the acquisition method
introduces a bias such that the
probability, $\p(i)$ that a paper is included in CiteSeer is a function
of number of authors of that paper. That is,
\begin{align}
\dblp(i) &= \alpha \cdot \all(i) \label{eq:one}\\
\citeseer(i) &= \p(i) \cdot \all(i) = p(i) \cdot \frac{\dblp(i)}{\alpha} \label{eq:dummy}
\end{align}

\subsubsection{CiteSeer Submission model}

Let $\beta \in (0,1)$ be the probability that an author submits a paper directly to CiteSeer then $p(i)=1-(1-\beta)^i$ where $ (1 - \beta)^{i}$ is the probability that none of the $i$ authors submit their paper to CiteSeer.

Substituting to (\ref{eq:dummy}) and re-arranging, we have
\begin{align}\label{eq:submissionmodel}
\ratio(i) = \frac{\dblp(i)}{\citeseer(i)} &= \frac{\alpha}{\left(1 - ( 1 - \beta )^{i}\right)}
\end{align}
It is clear from Equation~\ref{eq:submissionmodel} that as the number
of authors, $i$, increases, the ratio, $\ratio(i)$, tends to $\alpha$, i.e. we
expect that the number of $i$-authored papers in CiteSeer will
approach $\all(i)$ and thus from Equation~\ref{eq:one} the ratio tends
to $\alpha$. 
For single authored papers, i.e. $i=1$, we have that
$\ratio(1)=\frac{\alpha}{\beta}$ and since we know that DBLP has more
single-authored papers, it must be the case that $\beta<\alpha$.
More generally, we expect the ratio, $\ratio(i)$, to monotonically
decrease with the number of authors, $i$, reaching an asymptote of
$\alpha$ for large $i$.
This is approximately observed in
Figure~\ref{fig:authprob-dblpcsratio.pdf}, ignoring points for $n>30$
for which there is a scarcity of data.

In Figure \ref{fig:fit-a0-3-b0-15.pdf} we plot the proportion
$r(i)$ for numbers of authors $i$ where we have
at least 100 papers available. 
We fit Equation~\ref{eq:submissionmodel} to the data in 
Figure~\ref{fig:fit-a0-3-b0-15.pdf}\footnote{
Note that this is the same data as
Figure~\ref{fig:authprob-dblpcsratio.pdf} but restricted to $n<17$.
}. We see the fit is not perfect suggesting that this is not the only mechanism in play.

The value to which the data points are converging for high numbers of authors is $\alpha \approx 0.3$.  We have to take into account that we only used 71\% of DBLP papers and 57\% of CiteSeer papers in our analysis -- the papers that have both year and number of authors specified. Substituting $\alpha$ into \eqref{eq:adjust} we get the value of $\alpha' \approx 0.24$. If our model is correct, this would suggest that the DBLP database covers approximately 24\% of the entire Computer Science literature.

\begin{align}
\alpha'=\frac{\cdblp(i)}{\cciteseer(i)} &= \frac{0.57}{0.71} \cdot \frac{\dblp(i)}{\citeseer(i)} = 0.8 \cdot \alpha\label{eq:adjust}
\end{align}

\doublefig{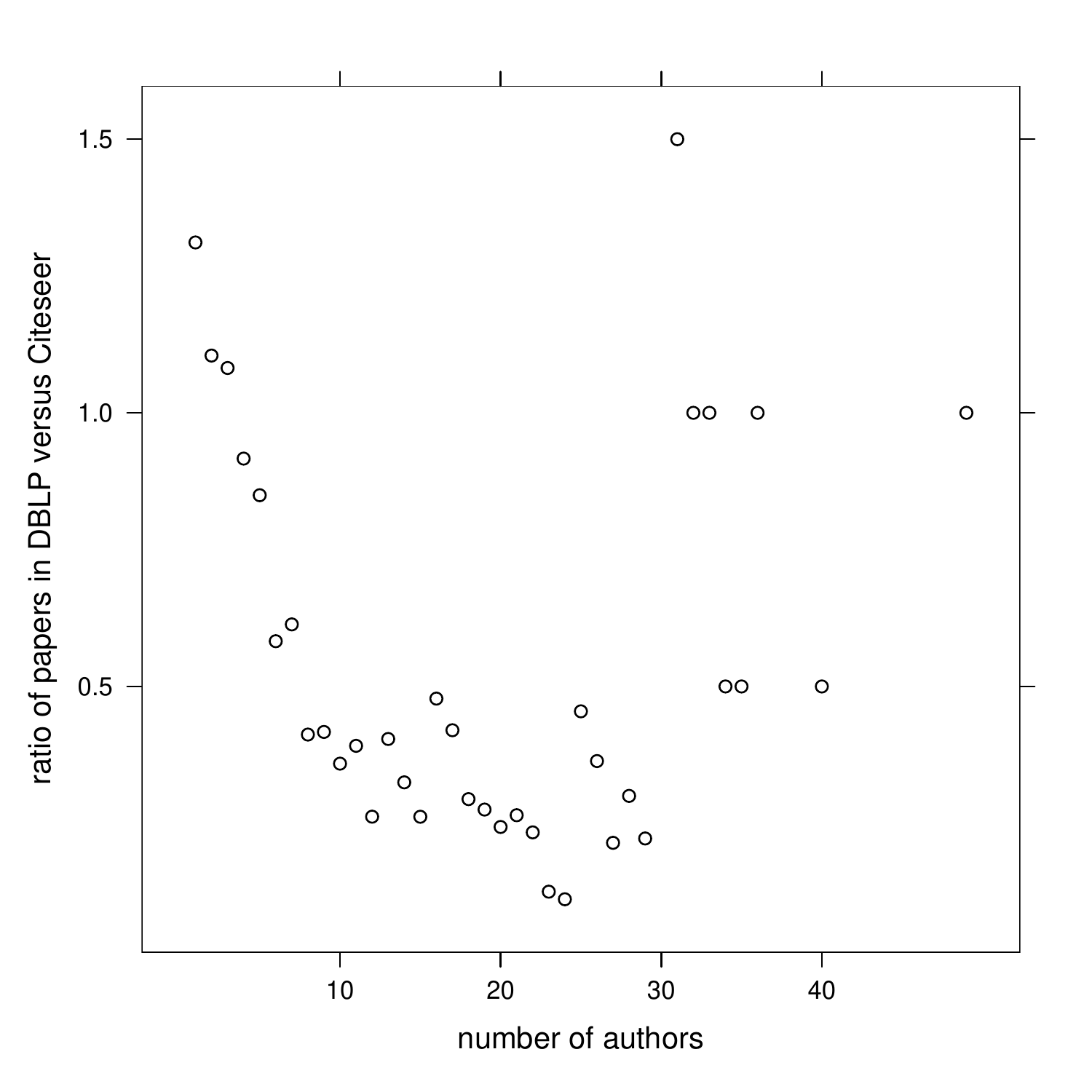}{Ratio of the number of authors in DBLP to CiteSeer as a function of the number of authors of a paper.}{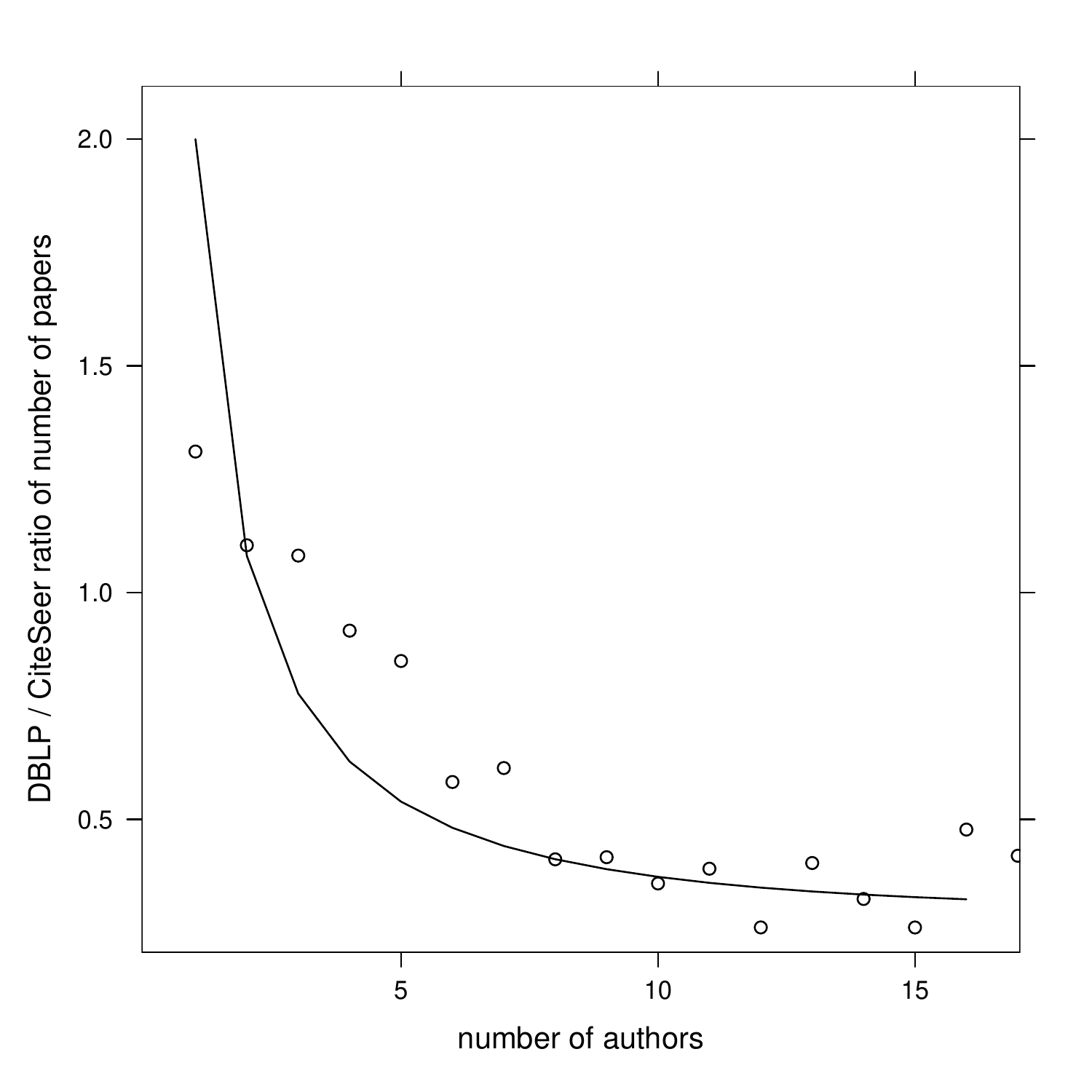}{Fit of model \eqref{eq:submissionmodel} for values of $\alpha=0.3$ and $\beta=0.15$ for numbers of authors where there are at least 100 documents in both datasets in total.}

\subsubsection{CiteSeer Crawler Model}

CiteSeer not only acquires papers based on direct submission by
authors, but also by a crawl of the Web.

To begin, let $\delta \in (0,1)$ be the probability that an author puts a
paper on a web site (homepage for example). Then the average number of
copies of an $n$-authored paper on the Web is $n \cdot \delta$. Let us
further assume that
the crawler finds each available on-line copy with a probability
$\gamma$. If $\pp(i,c)$ denotes the probability that there will be $c$
copies of an an $i$-authored paper published on-line, then we have:

\begin{center}
\begin{tabular}{c|ll}
authors     & pp(i,c)              & \\
\hline
$1$ & $\pp(1,1)=\delta$                           &  1 copy online\\
    & $\pp(1,0)=1-\delta$                         &  0 copies online\\
\\
$2$ & $\pp(2,2)=\delta^2$                         &  2 copies online\\
    & $\pp(2,1)=2\delta(1-\delta)$                &  1 copy online\\
    & $\pp(2,0)=(1-\delta)^2$                     &  0 copies online\\
\vdots \\
$n$ & $\pp(n,c)=\binom{n}{c} \delta^c(1-\delta)^{n-c}$ & $c$ copies online\\
& & of an\\
& & $n$-authored\\
& & paper
\end{tabular}
\end{center}

The probability, $\pf(c)$, of crawling a document with $c$ copies online, is
\begin{equation}
\pf(c) = 1 - ( 1 - \gamma )^{c}
\end{equation}
thus the probability that CiteSeer will crawl an $n$-authored
document, $\p(n)$ is 
\begin{align}
\p(n) &= \sum_{c=0}^n \pp(n,c)\pf(c)\notag\\
     &= \sum_{c=0}^n \pp(n,c)(1 - (1 - \gamma)^{c})\notag\\
     &= \sum_{c=0}^n \left(\binom{n}{c} \delta^c(1-\delta)^{n-c}\right) (1 - ( 1 - \gamma)^{c})\notag\\
     &= 1 - \sum_{c=0}^n \left(\binom{n}{c} \delta^c(1-\delta)^{n-c}\right) ( 1 - \gamma)^{c}\tag{sum of probabilities equals 1}\\
     &= 1 - \sum_{c=0}^n \left(\binom{n}{c}((1-\gamma) \delta)^c(1-\delta)^{n-c}\right)\notag\\
     &= 1 - (\delta(1-\gamma) + (1-\delta))^{n} \tag{from binomial theorem}\\
     &= 1 - (\delta-\gamma \delta + 1 -\delta)^{n}\notag\\
     &= 1 - (1 - \gamma \delta)^{n}
\label{eq:pncrawler}
\end{align}
where $ (1 - \gamma \delta)^{n}$ is the probability that no copy of
an $n$-author paper is found by CiteSeer.

Once again, if we substitute Equation~\eqref{eq:pncrawler} in
\ref{eq:dummy}, we have
\begin{align}\label{eq:crawlermodel}
\ratio(i)= \frac{\dblp(i)}{\citeseer(i)} &= \frac{\alpha}{\left(1 - ( 1 - \gamma \delta)^{i}\right)}
\end{align}
which is equivalent to the ``submission'' model of
Equation~\ref{eq:submissionmodel}. That is, both  models lead to
the same bias.

\section{Prior work}
\label{sec:previous}

There has been considerable work in the area of citation analysis and
a comprehensive review is outside of the scope of this paper. Broadly,
prior citation analysis has examined a wide variety of factors
including 
(i) the distribution of citation rates~\cite{redner@popular,lehmann@citation,tsallis@are,laherrere@stretched}, 
(ii) the variation in the distribution of citation rates across research fields
 and geographical regions~\cite{lehmann@citation,kim@comparative}, 
(iii) the geographic distribution of highly cited scientists~\cite{batty@citation,batty@geography} 
(iv) various indicators of the scientific performance of countries
\cite{may@scientific} 
(v) citation biases and miscitations~\cite{kotiaho@miscitation,kotiaho@unfamiliar,simkin@read}
(vi) collaboration networks~\cite{newman@structure}
(vii) distribution of references in papers~\cite{vazquez@statistics}, and
(viii) visualization and navigation~\cite{klink@browsing,cosley02referee}.

The number of citations is the most widely used measure of academic
performance and as such it influences decisions about distribution of
financial subsidies. The study of citation distributions helps us
understand the mechanics behind citations and objectively compare scientific performance.  

With regard to the distribution of citations,
Laherrere {\em et al}~\cite{laherrere@stretched} argued that a stretched
exponential\footnote{Stretched exponential distribution has the form $exp(-(x/w)^c)$} is suitable for modeling citation distributions as it is
based on multiplicative processes and does not imply an unlimited number
of authors. 
Redner~\cite{redner@popular} then analyzed the ISI and
Physical Review databases and showed that the number of citations of highly
cited papers follows a power-law.  Lehmann~\cite{lehmann@citation} attempted to fit both a power law and stretched
exponential to the citation distribution of 281,717 papers in the SPIRES~\cite{spires} database and showed it
is impossible to discriminate between the two models.

So far most of the research on citation distributions has come from the
Physics community.  Surprisingly little work has been done on computer
science papers. The ISI dataset contains computer science papers but
these were usually studied together with other disciplines despite the
fact that their dynamics may differ. The only work the authors are aware
of~\cite{newman@structure} is based on a small dataset (13000
papers) and was concerned with the distribution of the number of
collaborators. 

In the next section we examine the distribution of citations in both
the CiteSeer and DBLP datasets.

\subsection{Citation distributions for Computer Science}

Citation linking in DBLP was a one-time project performed as a part of the 'ACM SIGMOD Anthology' - a CD/DVD publication.  The citations were entered manually by students paid by ACM SIGMOD. As a result DBLP now contains a significant number of new papers that have not been included in this effort. To mitigate against this distortion, we limit ourselves in both datasets to papers that have been cited at least once (CiteSeer 100,059 papers, DBLP: 10,340 papers).

\begin{figure}[ht]
\begin{center}
\subfigure[atomic histogram]{\includegraphics[width=.48\textwidth]{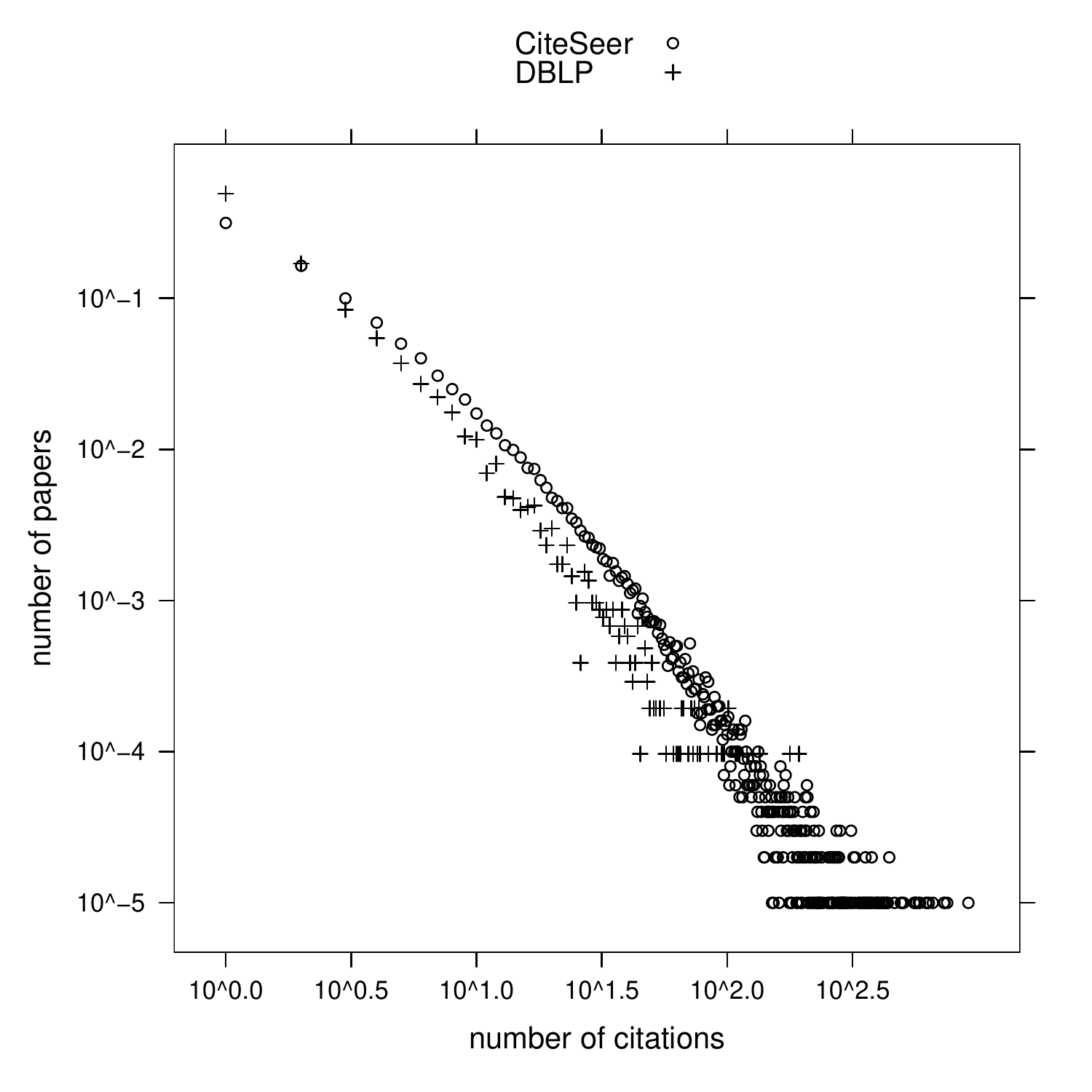}\label{fig:citprobhistogram-loglog.pdf}}
\subfigure[histograms after exponential binning]{\includegraphics[width=.48\textwidth]{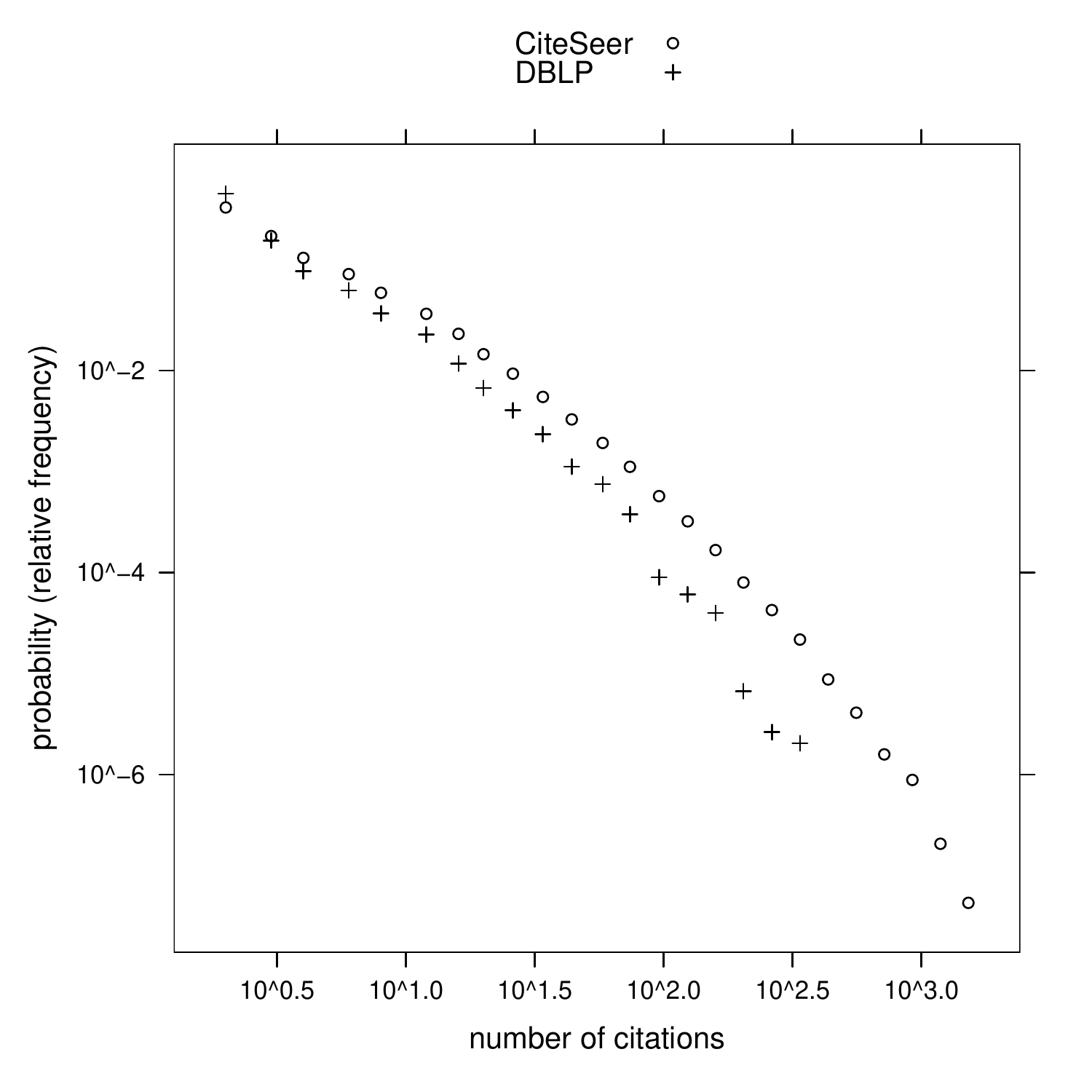}\label{fig:binnedprobhistogram.pdf}}
\caption{Probability histograms on double logarithmic scales for number of citations in the two datasets}
\end{center}
\end{figure}

Figure \ref{fig:citprobhistogram-loglog.pdf} compares citation
distributions in CiteSeer versus DBLP. We see that DBLP contains more
low cited papers than CiteSeer. We currently do not have an
explanation for this phenomenon. However, it may be related to Lawrence's~\cite{lawrence01online} observation that articles freely available online are more highly cited.

We use exponential binning (Figure~\ref{fig:binnedprobhistogram.pdf})
to estimate the parameters of the citation distribution in CiteSeer
and DBLP. Exponential binning is a technique where the data are
aggregated in exponentially increasing `bins'. In this manner we obtain
a higher number of samples in each bin, which reduces the noise in the data.

The slopes in Table \ref{tab:slopes} correspond to linear
interpolation of exponentially binned data as displayed in
Figure~\ref{fig:binnedprobhistogram.pdf}. Higher slopes in our
datasets indicate a more uneven distribution of citations. The papers in
each dataset have been divided into two groups -- papers with more
than and less than 50 citations.%\footnote{50 citations have been used by Lehmann and should be replaced by some other value as each dataset contains a different number of citations. Nevertheless this number happens to divide both datasets well.}

For both datasets we obtain parameters bigger in absolute value than
Leh\-mann~\cite{lehmann@citation} derived for Physics. This means that
highly cited papers acquire a larger share of citations in Computer
Science than in Physics.%\footnote{ Lehmann actually did exponential and stretched exponential fitting.  }
However, there is also a significant difference between CiteSeer and DBLP.

\begin{table}[ht]
\begin{center}
\begin{tabular}{c|c|c|c}
number                 & \multicolumn{3}{c}{slope}\\
of citations & Lehmann & CiteSeer & DBLP      \\
\hline
$<50$ & -1.29 & -1.504 & -1.876 \\
$>50$ & -2.32 & -3.074 & -3.509 \\
\end{tabular}
\caption{Slopes for Figure \ref{fig:binnedprobhistogram.pdf} 
  representing the parameter of the corresponding power-laws.
\label{tab:slopes}}
\end{center}
\end{table}

\section{Conclusions}

This paper compared two popular online science citation databases,
DBLP and CiteSeer, which have very different methods of data
acquisition. We showed that autonomous acquisition by web crawling,
(CiteSeer), introduces a significant bias against 
papers with low number of authors (less than 4). Single author papers appear to be disadvantaged with regard to the CiteSeer acquisition method. As such, single authors, (who care) will need more actively submit their papers to CiteSeer if this bias is to be reduced.

We attempted to model this bias 
by constructing two probabilistic models for paper
acquisition in CiteSeer. The first model assumes the probability that
a paper will be submitted is proportional to the number of authors of
the paper. The second model assumes that the probability of crawling a
paper is proportional to the number of online copies of the paper and
that the number of online copies is again proportional to the number
of authors.  Both models are equivalent and permit us to estimate that
the coverage of DBLP is approximately 24\% of the entire Computer Science
literature.

We then examined the citation distributions for both CiteSeer and DBLP
and observed that CiteSeer has a fewer number of low-cited papers. The
citation distributions were
compared with prior work by Lehmann~\cite{lehmann@citation}, who examined datasets from the Physics community. While the CiteSeer and DBLP
distributions are different, both datasets exhibit steeper slopes than
SPIRES HEP dataset, indicating that highly cited papers in Computer Science
receive a larger citation share than in Physics.

\section{Acknowledgments}

The authors thank Michael Ley for his assistance in understanding DBLP.

\end{document}